\documentclass[aps,prc,showpacs,amssymb,amsmath,amsfonts,nofootinbib,floatfix]{revtex4-1}
\usepackage{graphicx}

\makeatletter
\def\p@subsection{}
\def\p@subsubsection{}
\makeatother
\usepackage{natbib}
\usepackage[usenames]{color}
\usepackage{epsfig}
\usepackage{epstopdf}
\usepackage{amssymb,amsmath}
\usepackage{amsmath}
\usepackage{subfig}
\usepackage{epstopdf}
\usepackage{enumitem}
\usepackage{bbm}

\definecolor{grey}{rgb}{0.9,0.9,0.9}
\definecolor{black}{rgb}{0,0,0}

\newcommand{\di}{i}

\newcommand{\be}{\begin{eqnarray}}
\newcommand{\ee}{\end{eqnarray}}
\newcommand{\bc}{\begin{center}}
\newcommand{\ec}{\end{center}}
\newcommand{\beq}{\begin{eqnarray}}
\newcommand{\eea}{\end{eqnarray}}

\begin{document}

\title{A connection between angle-dependent phase ambiguities
       and the uniqueness of the partial-wave decomposition }

\author{ A. \v{S}varc\,$^{1}$ \\
 Y. Wunderlich\,$^2$, H. Osmanovi\'{c}\,$^3$ \\
  M. Had\v{z}imehmedovi\'{c}$\,^3$, R. Omerovi\'{c}$\,^3$, J. Stahov$\,^3$, \\  V. Kashevarov$\,^4$, K. Nikonov$\,^4$, M. Ostrick$\,^4$, L. Tiator$\,^4$  \\
R. Workman$^5$   }
\email{Corresponding author: alfred.svarc@irb.hr}
\affiliation{$\,^1$ Rudjer Bo\v{s}kovi\'{c} Institute, Bijeni\v{c}ka cesta 54,
                 P.O. Box 180, 10002 Zagreb, Croatia}
\affiliation{$\,^2$ Helmholtz-Institut f\"{u}r Strahlen- und Kernphysik der Universit\"{o}t Bonn, Nussallee 14-16, 53115 Bonn, Germany }
\affiliation{$\,^3$ University of Tuzla, Faculty of Natural Sciences and Mathematics, Univerzitetska 4, 75000 Tuzla, Bosnia and Herzegovina}
\affiliation{$\,^4$ Institut f\"{u}r Kernphysik, Universit\"{a}t Mainz, D-55099 Mainz, Germany}
\affiliation{$\,^5$ Data Analysis Center at the Institute for Nuclear Studies, Department of Physics, The George Washington University, Washington, D.C. 20052, USA}

\vspace{5cm}
\date{\today}

\begin{abstract}
Unconstrained partial-wave amplitudes, obtained at discrete energies from fits to complete sets of eight independent observables, may be used to reconstruct reaction amplitudes. These partial-wave amplitudes do not vary smoothly with energy and are in principle non-unique. We demonstrate how this behavior can be ascribed to the continuum ambiguity. Starting from the spinless scattering case, we show how an unknown overall phase, depending on energy and angle, mixes the structures seen in the associated partial-wave amplitudes. This process is illustrated using a simple toy model. We then apply these principles to pseudo-scalar meson photo-production, showing how the above effect can be removed through a phase rotation, allowing a consistent comparison with model amplitudes. The effect of this phase ambiguity is also considered for Legendre expansions of experimental observables.

\end{abstract}

\maketitle
\section{Introduction}

Partial-wave analysis, a textbook method to identify resonances and determine them quantum numbers, is a standard procedure used to analyze a wide class of experimental data
(see for instance ref. \cite{MartinSpearman}).
This paper deals with a very general interplay of  partial-wave decomposition and invariance with respect to a general symmetry
of energy- and angle-dependent phase rotation, and is therefore of wide
interest for all scientists who apply the method of partial-wave decomposition
in their analysis of data.
 This phase invariance causes a critical, through rarely addressed, problem of
non-uniqueness in partial-wave decomposition, which has a profound impact on
the general problem of resonance identification.

In the quest for a unique set of amplitudes,
experimental programs have attempted to measure complete sets of observables needed to perform the unambiguous  reaction-amplitude reconstruction.  On the other hand, it has also been well known that single-channel physical observables remain invariant with respect to a general energy- and angle-dependent phase rotation which turns out to introduce the non-uniqueness into the amplitude reconstruction.
This invariance is the so-called continuum ambiguity, and has been extensively analyzed in mid-70s through
the mid-80s \cite{Atk73,Bow75,Atk85}.  The interconnection of the two has never been discussed as  most studies were made in the context of $\pi N$ elastic scattering where the
optical theorem and an application of elastic unitarity practically eliminate the continuum ambiguity as a source of non-uniqueness.  In the inelastic domain, up to now, the main attention has been paid only to handling angle independent phase rotations on the level of partial waves where the analytic structure  remains untouched \cite{bnga,Tiator:2011pw,Sandorfi2011}.  We show that, contrary to well-behaved energy dependent invariance, angle dependent phase rotation mixes partial waves, and changes their analytic structure causing partial-wave decomposition to be non-unique. This mixing directly influences the present identification of resonance quantum numbers. The discussion of angle dependent phase rotation was always scarce, and to our knowledge, mentioned only "in passing" in old ref.~\cite{Oma81}, and very recently, but in different context, also in \cite{Gibs2008}.
Some other related studies on phase ambiguities were performed \cite{Dean,Keaton}, but concentrated on cases without a complete set
of experimental data. Programmatic studies of photo-production experiments at Jefferson Lab, Mainz and Bonn are now producing the data required to do complete experiments, in terms of either helicity amplitudes or multipoles, motivating a reexamination of
the ambiguities associated with multipole analyses.

 It is well known that the importance of phase invariance is very different for elastic,  slightly inelastic, and strongly inelastic processes, and in this paper  we are primarily focused on strongly inelastic, photo-production reactions.  We show that, contrary to well-behaved energy-dependent phase rotations, angle-dependent phase rotations
mix partial waves, changing their analytic structure and producing partial-wave decompositions which are
non-unique. This directly influences the present identification of resonance quantum numbers.   We then explore the effect of angle-dependent rotations in a
case of practical interest, the single-energy (SE) partial-wave analysis (PWA)  of a complete set of eta photo-production numeric data generated from a known model. Here the extracted multipoles can be directly compared to those of the underlying model. We confirm that the obtained result is non-unique.  As a solution of this problem we offer a method how a partial-wave decomposition can be made up-to-a-phase unique by replacing the undetermined and discontinuous energy and angle dependent phase with an arbitrary but continuous chosen value. This is by no means a restriction of generality as all other solutions with different amplitude phases can be obtained directly from the obtained solution by a phase rotation on the level of reaction amplitudes, changing the arbitrary to any particularly
 given phase. All solutions remain equivalent as far as fitting the data is concerned. For the true quantum number identification, however, one still has to identify the unknown continuum ambiguity phase because each phase gives its own set of quantum numbers. In fully unconstrained single-channel analyses, quantum numbers remain undetermined.
\section{Formalism}
Let us recall that observables in single-channel reactions are given as a sum of products involving one amplitude
(helicity, transversity, ... ) with the complex conjugate of another, so that the general form of
any observable is given as ${\cal O}=f˙(H_k \cdot H_l^*)$, where $f$ is a known, well-defined real function.
The direct consequence is that any observable is invariant with respect to the following simultaneous
phase transformation of all amplitudes:
\be
H_k (W,\theta) \rightarrow  \tilde{H_k} (W,\theta)& = & e^{\, i \, \,  \phi(W,\theta) } \cdot H_k (W,\theta) \label{eq:ContAmbGeneralTrafo} \\
 {\rm for \, \, all} \, \, k & = & 1,\cdots,n \nonumber
\ee
\noindent
where n is the number of spin degrees of freedom (n=1 for the 1-dim toy model, n=2 for pi-N scattering
and n=4 for pseudoscalar meson photo-production), and $\phi(W,\theta)$ is an arbitrary, real function which is the same for all contributing amplitudes.

Without any further physics constraints, this real function $\phi(W,\theta)$ is free, and there exist an infinite number of equivalent solutions which give exactly the same set of observables. However, the unitarity  condition  plays an important role in constraining this invariance. Defining the S- and T-matrix as abstract operators in a Hilbert-space of scattering states, i.e. $\hat{S} = \mathbbm{1} + 2 i \hat{T}$, unitarity has to hold for the S-matrix $\hat{S}^{\dagger} \hat{S} = \mathbbm{1}$ and it leads to the optical theorem in operator-notation (cf. reference \cite{Peskin} for the ensuing discussion):
\begin{equation}
 \frac{1}{2 i} \left( \hat{T} - \hat{T}^{\dagger} \right) = \hat{T}^{\dagger} \hat{T} \mathrm{.} \label{eq:OptTheoremInOperatorNotation}
\end{equation}
In order to get a relation among actual amplitudes out of this operator-equation, the latter has to be sandwiched between initial and final states $\left< f \right| \left( \ldots \right) \left| i \right>$. The T-matrix amplitude is then the complex function $T_{fi} \equiv \left< f \right| \hat{T} \left| i \right>$ depending on a minimal set of external invariants for the reaction. For a $2 \rightarrow 2$-reaction, these would be just the Mandelstam variables $(s,t)$. In most cases, the invariant scattering amplitude is defined by factoring out a $\delta$-function $ \left< a \right| \hat{T} \left| b \right> = T_{ab} = (2 \pi)^{4} \delta^{(4)} \left( P_{b} - \tilde{P}_{a} \right) A_{ab}$, with $P_{b}$ and $\tilde{P}_{a}$ the total $4$-momenta in states $\left| b \right>$ and $\left| a \right>$, respectively. \newline
Inserting a complete set of kinematically allowed intermediate states in the operator product of the right hand side of (\ref{eq:OptTheoremInOperatorNotation}), one gets:
\begin{equation}
 \frac{1}{2 i} \left( A_{fi} - A_{if}^{\ast} \right) = \sum_{j} \int d \Phi_{j} A_{jf}^{\ast} A_{ji} \mathrm{,} \label{eq:UnitarityEquationFullAmplitude}
\end{equation}
where $\left< a \right| \hat{T}^{\dagger} \left| b \right> = \left< b \right| \hat{T} \left| a \right>^{\ast}$ was used and $d \Phi_{j}$ the Lorentz-invariant phase-space differential for intermediate state $\left| j \right>$. We now specify this for the special case of a fully elastic reaction, and we assume that interactions are time reversal invariant and only scalar particles participate, with two particles in the initial and final state. Then, the left-hand-side of (\ref{eq:UnitarityEquationFullAmplitude}) simplifies to an imaginary part and one obtains
\begin{equation}
 \mathrm{Im} \left[ A^{\text{el}}_{fi} \right] = \int d \Phi_{m} \left(A^{\text{el}}_{mf}\right)^{\ast} A^{\text{el}}_{mi} \mathrm{.} \label{eq:UnitarityElastic}
\end{equation}
The only accessible intermediate state is called $\left| m \right>$ and it contains the same particles as $\left| i \right>$ and $\left| f \right>$, but kinematics may differ. If we now abbreviate the elastic amplitude as $  A^{\text{el}}_{fi} \equiv A^{\text{el}}$ and expand this amplitude into partial waves, it is seen that the integral constraint (\ref{eq:UnitarityElastic}) for the full amplitude nicely decouples for the individual partial waves \cite{Gribov}:
\begin{equation}
 \mathrm{Im} \left[ A^{\text{el}}_{\ell} \right] = \left| A^{\text{el}}_{\ell} \right|^{2} \mathrm{.} \label{eq:UnitarityPartialWaves}
\end{equation}
We can rewrite this condition as $U^{\text{el}} := \mathrm{Im} \left[ A^{\text{el}}_{\ell} \right] - \left| A^{\text{el}}_{\ell} \right|^{2} = 0$ and it is generally solved by $A^{\text{el}}_{\ell} \propto e^{i \delta_{\ell}} \sin \delta_{\ell}$ with a real phase-shift $\delta_{\ell}$. This  equality condition is called elastic unitarity.
So, in elastic region the unitarity is an equality, and an additional equation relating real and imaginary part of the T-matrix is introduced. The function $\phi(W,\theta)$, in principle, becomes fully determined. However, Crichton \cite{Chri66}  has  for the special case of scalar field  within the context of a truncated partial-wave analysis up to $D$-waves, found that elastic unitarity (\ref{eq:UnitarityPartialWaves}) combined with data fixes the amplitude only up to one non-trivial discrete ambiguity, so some discrete ambiguities are left. Later, in \cite{ItzyksonMartin,AtkinsonEtAl,Bow75}, strong hints were found that one non-trivial discrete ambiguity would still be allowed by elastic unitarity independently of the truncation order $\ell$  and the dimensionality of the process, but the issue has never been fully settled. Anyway, the reduction of continuum ambiguity in elastic region is almost complete.
\\ \\ \indent
 Stepping into inelastic region of elastic scattering, equality becomes inequality {$U^{\text{el}} \geq 0$}, and the additional equation is lost. This inequality condition is called inelastic unitarity.  However, the T-matrix  of elastic scattering in the near-threshold inelastic region   must still satisfy inelastic unitarity, and despite being only an inequality, it also introduces serious restrictions upon possible choices of rotating function $\phi(W,\theta)$. This effect has been extensively discussed in refs.~\cite{Gibs1978,Gibs2008} for elastic scattering in near-threshold inelastic region, and it was shown that inelastic unitarity practically eliminates all solutions obtained by the angle dependent phase rotation of the initial one found in a unitarity constrained fit. However, for inelastic processes  the situation changes significantly. The form of inelastic unitarity is different, and inelastic unitarity  {$U^{\text{el}}$} for elastic scattering is replaced with inelastic unitarity for inelastic processes {$U^{\text{inel}} := \mathrm{Im} \left[ A^{\text{el}}_{\ell} \right] - \left| A^{\text{el}}_{\ell} \right|^{2} - \left| A^{\text{inel}}_{\ell} \right|^{2} \geq 0$}, see ref. \cite{Selleri1968}.
The inelastic unitarity of inelastic channels, therefore, demands that the absolute value of inelastic T-matrix should be smaller than the available ``room" which is left by the elastic unitarity {$\mathrm{Im} \left[ A^{\text{el}}_{\ell} \right] - \left| A^{\text{el}}_{\ell} \right|^{2} \geq 0$}. So, we cannot speak about inelastic unitarity without specifying the elastic part. As we are primarily interested in photo-production processes, this condition is in this case trivially satisfied. First, the dominance of elastic cross sections over inelastic electromagnetic ones suggests that unitarity will not strongly constrain the
associated photoproduction amplitudes. In addition, the electromagnetic amplitudes themselves (cross section is related to the S-matrix by
flux factors) are also significantly smaller than the elastic and other inelastic counterparts, so elastic channel unitarity does leave ample amount of ``room", and inelastic unitarity is practically never violated for any phase rotation.
\\ \\ \indent
Having completed the analysis of conditions when the general invariance given by Eq.~(\ref{eq:ContAmbGeneralTrafo}) is valid,  we focus on  resonance properties of amplitudes $ H_k (W,\theta) $ which are usually the goal of such studies. These are usually identified with poles of the partial-wave (or multipole) amplitudes, we must analyze the influence of the continuum ambiguity not
upon helicity or transversity amplitudes, but upon their partial-wave decompositions.
To simplify the study we introduce partial waves in a simplified version than those found in
ref. \cite{Tiator:2011pw}:
\be \label{Eq:PW}
A (W,\theta) & = & \sum^{\infty}_{\ell=0} (2 \ell + 1) A_\ell(W) P_{\ell}(\cos\theta)
\ee
\noindent
where $A(W,\theta)$ is a generic notation for any amplitude $H_k(W,\theta)$, $k=1, \cdots n$.
The complete set of observables remains unchanged when we make the following transformation:
\be \label{Eq:PWrot}
A (W,\theta) \rightarrow \tilde{A} (W,\theta)& = & e^{\, \di \, \,  \phi(W,\theta) }
 \hspace*{5pt} \times \hspace*{5pt}   \sum^{\infty}_{\ell=0} (2 \ell + 1) A_\ell(W) P_{\ell}(\cos\theta) \nonumber \\
\tilde{A} (W,\theta) & = &  \sum^{\infty}_{\ell=0} (2 \ell + 1) \tilde{A}_\ell(W)P_{\ell}(\cos\theta)
\ee
\noindent
We are interested in rotated partial wave amplitudes $\tilde{A}_\ell(W)$, defined by Eq.(\ref{Eq:PWrot}),
and are free to introduce the Legendre decomposition of an exponential function as:
\be \label{Eq:Phaseexpansion}
e^{\, \di \, \,  \phi(W,\theta) } &=& \sum^{\infty}_{\ell=0} L_\ell(W)  P_{\ell}(\cos\theta).
\ee
After some manipulation of the product
$P_\ell(x) P_k(x)$
(see refs.~\cite{Dougall1952,Wunderlich2017}~for details of the summation rearrangement) we obtain:
\be \label{Eq:mixing}
\tilde{A}_\ell(W) &=& \sum_{\ell'=0}^{\infty} L_{\ell'}(W) \, \, \, \cdot \sum_{m=|\ell'-\ell|}^{\ell'+\ell}\langle \ell',0;\ell,0|m,0 \rangle ^2\, \, A_{m}(W) \nonumber \\
\ee
where $\langle \ell',0;\ell,0|m,0 \rangle $ is a standard Clebsch-Gordan coefficient.  A similar relation was also derived in ref.~\cite{Gibs2008}.
\\ \indent
To get a better insight into the mechanism of multipole mixing, let us expand Eq.~(\ref{Eq:mixing}) in terms of phase-rotation Legendre coefficients $L_{\ell'}(W)$: \\
\begin{align} \label{Eq:mixing-expanded}
 \tilde{A}_{0} (W) &= {L_{0} (W) A_{0} (W) } + L_{1} (W) A_{1} (W) + L_{2} (W) A_{2} (W) + \ldots  \mathrm{,} \\
 \tilde{A}_{1} (W) &= {L_{0} (W) A_{1} (W) } + L_{1} (W) \left[\frac{1}{3}A_{0} (W) + \frac{2}{3} A_{2} (W) \right] + L_{2} (W) \left[\frac{2}{5} A_{1} (W) + \frac{3}{5} A_{3} (W) \right] + \ldots \mathrm{,} \nonumber \\
 \tilde{A}_{2} (W) &= {L_{0} (W) A_{2} (W) } + L_{1} (W) \left[\frac{2}{5} A_{1} (W) +\frac{3}{5} A_{3} (W) \right] + L_{2} (W) \left[\frac{1}{5} A_{0} (W) + \frac{2}{7} A_{2} (W) + \frac{18}{35} A_{4} (W) \right] + \ldots \mathrm{.}  \nonumber \\
 \vdots \nonumber
\end{align}
The consequence of Eqs.~(\ref{Eq:mixing})~and~(\ref{Eq:mixing-expanded}) is that angular-dependent phase rotations
mix multipoles.
\\ \\ \noindent
\underline{\emph{Conclusion:}}
\\ \\ \indent
Without fixing the free continuum ambiguity phase $\phi(W,\theta)$,
the partial-wave decomposition $A_\ell (W)$ defined in Eq.~(\ref{Eq:PW}) is non-unique. Partial waves get mixed, and  identification of resonance quantum numbers might be changed.
To compare different partial-wave analyses, it is  essential to match the continuum ambiguity phase;
otherwise the mixing of multipoles is yet another, uncontrolled, source of systematic errors. Observe that this phase rotation does not create new pole positions, but just reshuffles the existing ones among several partial waves.

\section{Toy model:}

\begin{figure*}[h]
\begin{center}
\includegraphics[width=0.85\textwidth]{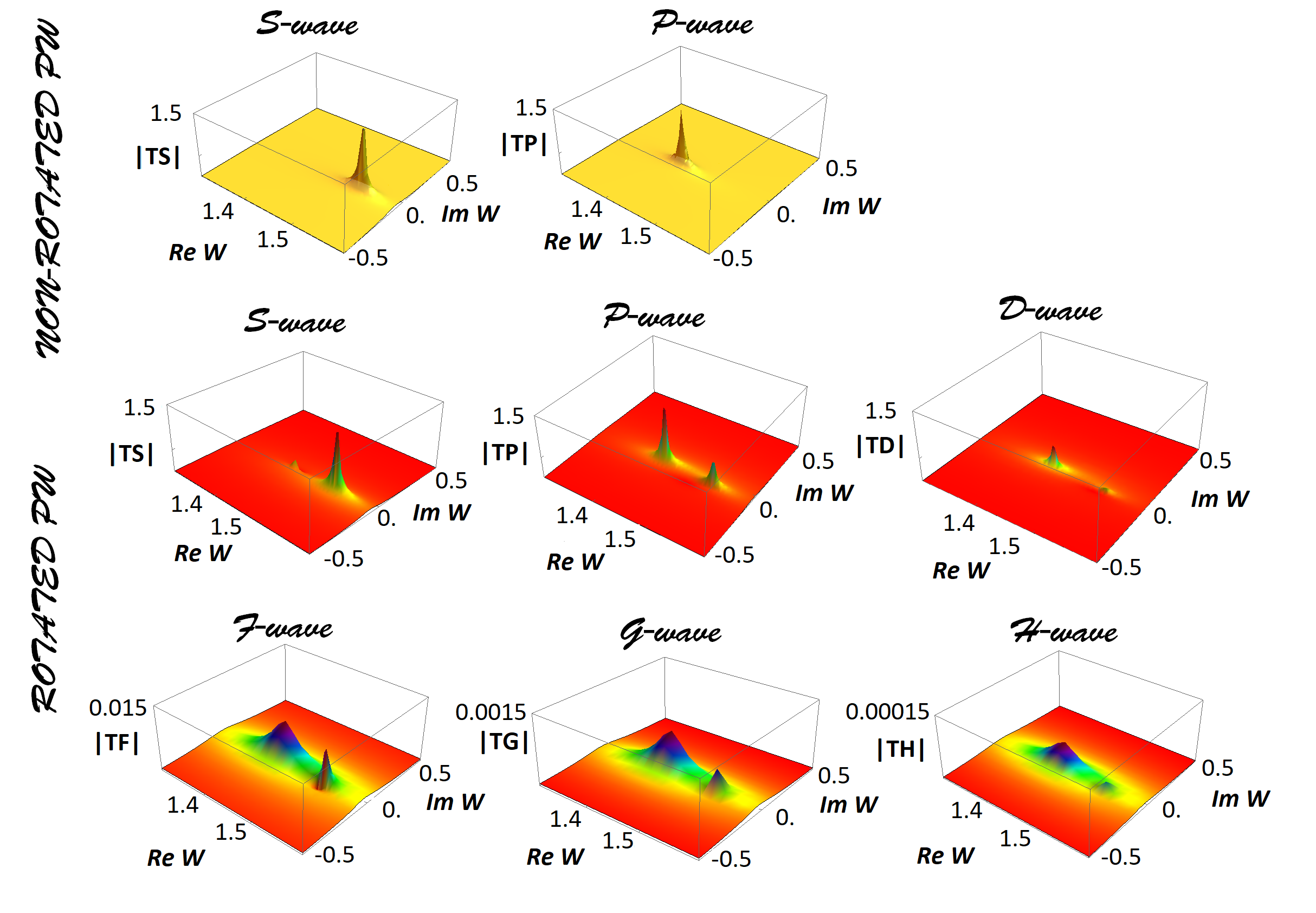}\vspace{-3mm}
\caption{\label{Toy model}(Color online) Toy model poles for linear rotation. }
\end{center}
\end{figure*}

To better illustrate the effect of mixing partial waves in the photo-production channel, we construct a simple toy model consisting of two partial waves,
with one resonance per partial wave.
\be
A(W,\theta) &=& T_S(W) + x \, \, T_P (W) \nonumber \\ \nonumber \\
T_{S,P}(W) &=&  \frac{a_{S,P}}{M_{S,P} - \di \, \Gamma_{S,P}/2-W} \\
     x & = & \cos \theta
\ee
where
\begin{center}
\begin{tabular}{ccc}
$a_S = (0.005 + \di \, 0.004$) GeV ; \hspace*{0.7cm}    & $M_S = 1.535 $ GeV; \hspace*{0.7cm} &  $\Gamma_S = 0.15$ GeV  \\
$a_P = (0.004 + \di \, 0.003$)  GeV ; \hspace*{0.7cm}   & $M_P = 1.440 $ GeV; \hspace*{0.7cm} &  $\Gamma_P = 0.10$  GeV.\\ \\
\end{tabular}
\end{center}
We take a very simple linear rotation acting upon the full amplitude $A(W,\theta)$:
\be \label{Toymodelrotation}
\mathcal{R}(x) &=& e^{ \di \, (2. + 0.5 \, x)},
\ee
and compare the partial-wave decompositions of the non-rotated and rotated amplitudes.
In Fig.~\ref{Toy model} we show the result.

We have already discussed the constraining role of inelastic unitarity in phase rotations of elastic scattering in the near-threshold inelastic region, therefore we construct our Toy model in such a way that inelastic unitarity is manifestly never violated.   Our Toy model is meant to describe processes in the photo-production channel. It is strongly inelastic, residues are very small, and the typical contribution to the inelastic unitarity relation $U^{inel}$ for both, non-rotated and rotated amplitudes is small compared to the available space left by the elastic channel. If one calculates the $|T_S(W=M_S)|^2$ and $|T_P(W=M_P)|^2$ for non-rotated amplitudes one obtains 0.0073 and 0.01 respectively, which is negligible compared to the space typically available from elastic unitarity at these energies, which is of the order of~0.1. For rotated amplitudes, contributions to the inelastic unitarity are similar, and even smaller.
 \\ \\
On the basis of Eqs.~(\ref{Eq:mixing})~and~(\ref{Eq:mixing-expanded}) applied to the toy model,
we observe several important features of the angle-dependent phase rotations:
\begin{enumerate}
\item The Legendre expansion of the angle-dependent phase rotation, given by
Eq.~(\ref{Eq:Phaseexpansion}), has, in principle, an infinite number of terms.
However, in cases of simpler rotations, higher order terms rapidly become very small. In this case the partial-wave mixing is limited only to the neighbouring
partial waves (see Eq.~(\ref{Eq:mixing-expanded})).\footnote{For more complicated rotations, the ones where associated Legendre coefficients
extend to infinity, the situation gets more unstable as the resonance with the very high angular momentum can be
intermixed even into usually dominating S-wave. So, association of quantum numbers becomes even more questionable.}
\item The lowest rotation coefficient $L_0(W)$ transforms  $A_\ell(W)$ to  $\tilde{A}_\ell(W)$.
        Hence, for angle-independent rotations only the 0-th order Legendre coefficient survives, and
such a rotation does not mix multipoles.
\item Even the simplest quickly converging rotations, with only a few Legendre coefficients
significantly greater than zero, will mix multipoles. The degree of mixing rises with the complexity of
the angular rotation.
\item As explicitly seen in Eq.~(\ref{Eq:mixing-expanded}) under the influence of any phase rotation  only existing poles gets redistributed. So, through the influence of the angle-dependent phase rotation, no new pole positions are created,
but resonances with the same pole positions appear in different partial waves making the quantum number identification difficult. The only way out is to identify the genuine
continuum ambiguity phase where the partial-wave matrix gets diagonal, and only one resonance is associated with one partial wave.\footnote{We can understand all partial waves as being a matrix elements of the partial-wave-operator defined on the space of Legendre polynomials which form an orthonormal basis. Without knowing the genuine continuum ambiguity phase that operator is non-diagonal, and same resonances appear in different matrix-element positions hence having different quantum numbers. However, we can always diagonalize the partial-wave-operator (find such a basis of Legendre polynomials where this operator is diagonal), and in this base only one partial wave is associated with one Legendre polynomial defining the true resonance quantum numbers. Let us add that this new base where the partial-wave operator is diagonal corresponds to the true continuum ambiguity phase.  } \\ \\ \indent
\end{enumerate}
Another interesting effect can be observed. The toy-model $A(W,\theta)$ truncated at $L=1$
leads to 'data' for the differential cross section whose Legendre-expansion
\begin{equation}
\sigma_{0} (W,\theta) = \left|A(W,\theta)\right|^{2} = \sum_{n=0}^{2} a_{n} (W) P_n (\cos \theta) \mathrm{,} \label{eq:DCSLegendreExpansion}
\end{equation}
would be precisely truncated at $2L=2\ast1=2$, i.e. all Legendre coefficients up
to and including $a_{2}$ would be non-vanishing, with all higher coefficients,
i.e. $a_{3}$ and above, being  precisely zero.
If one then defined a new amplitude, with the angle-dependent phase (\ref{Toymodelrotation})
multiplied into the original
truncated amplitude, we would end up with an, in principle, infinite partial-wave expansion
for the rotated amplitude (which, however, converges quite well
for the higher $\ell$).
In fact, the rotation (\ref{Toymodelrotation}) itself turns out to
be truncated at $L^{\prime}=5$ to a very good approximation, since all of its
Legendre coefficients $L^{\prime}{\geq6}$ have moduli within the range of $10^{-7}$ or less.

Now, one might naively expect the Legendre-expansion of the data
(\ref{eq:DCSLegendreExpansion}) to show non-vanishing Legendre-coefficients up to and including
$2(L + L^{\prime}) = 2*(1 + 5) = 12$, once the rotated amplitude has been inserted.
This, however, is is not what happens. Since the phase (\ref{Toymodelrotation}) has
modulus $1$ for all $x$, it will leave the cross section invariant. The data simply do not change.
Since the original data are truncated at $2\ast1 = 2$, by construction, the rotated data will be as well.
However, the rotated amplitude itself
is manifestly not truncated. Furthermore, higher Legendre-coefficients $a_{\, \ell \geq 3}$
are bilinear hermitean forms, depending on the rotated partial waves.
All this implies that a cancellation-effect has to set all 'rotated' Legendre coefficients
$a_{3}$,...,$a_{12}$ to zero. The rotation (\ref{Toymodelrotation}) has generated higher partial waves in such a
'finely tuned' way that this is indeed possible.

 \section{ Numeric data}
Having explored a simple toy model, we next consider a more realistic analysis of  numeric data data generated  with machine precision
from an existing model for this chosen reaction.  Minimization is done with statistical weight equal to one ($w = 1$). Just for the convenience of the reader let us summarize the essence of numeric data method.

Using numeric data for testing new procedures is a textbook method, commonly used in our field (see for instance ref~\cite{Numeric data}).  In general we use it to test whether a certain procedure is correct by applying it to a set of data artificially created from a known source, and for which we know the exact answer in advance.  In this paper  we use it to test whether phase rotation can be used to  eliminate the discontinuity  of unconstrained  SE PWA. We construct a complete set of observables out of  theoretical model-partial-waves, so we know in advance what should we get if the recommended method is correct. Then we perform the procedure, and if we reproduce the input-partial-waves, we know that our method is valid. Observe that the source of the generating model is unimportant as long as our procedure is consistently implemented,  so we shall not discuss the features of the model used.

We perform unconstrained, $L_{max}=5$ truncated single-energy analyses on a
complete set of observables for $\eta$ photoproduction given in the form of numeric data created using the
ETA-MAID15a model
\cite{MAID15a}: $d\sigma /d\Omega$, $ \Sigma \, d\sigma /d\Omega$, $ T \, d\sigma /d\Omega$, $ F \, d\sigma /d\Omega$, $ G \, d\sigma /d\Omega$, $ P \, d\sigma /d\Omega$, $ C_{x'} \, d\sigma /d\Omega$, and $ O_{x'} \,
d\sigma /d\Omega$.   All higher multipoles are put to zero. The unconstrained fitting procedure  with statistical weight equal to one ($w=1)$ turns out to be non-unique and  discontinuous. First, all solutions are discontinuous. Second, depending on the choice of initial parameters in the fit, we obtain numerous discontinuous, but with respect to $\chi^2$, equivalent solutions. We demonstrate this in forthcoming figures.

\begin{figure}[h!]
\begin{center}
\includegraphics[width=0.45\textwidth]{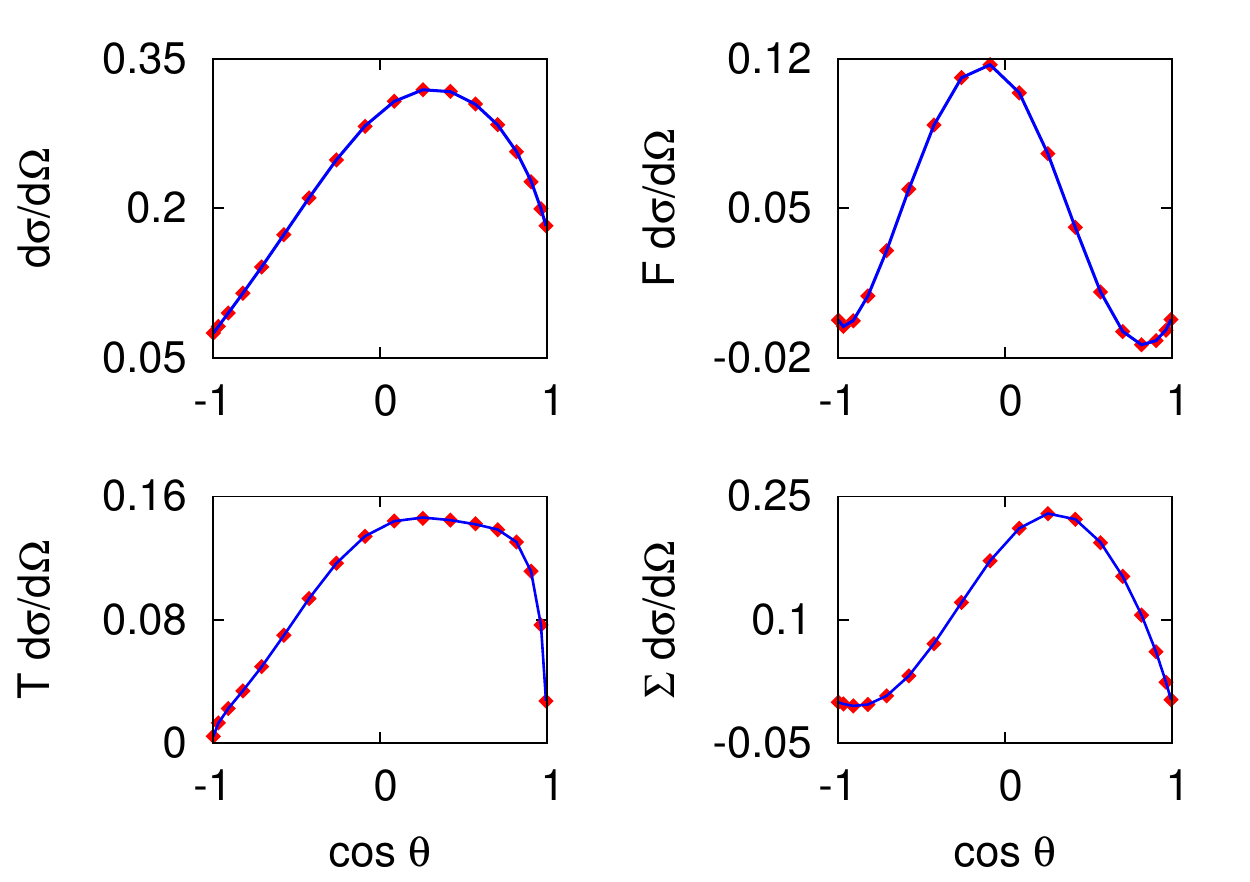}
\includegraphics[width=0.45\textwidth]{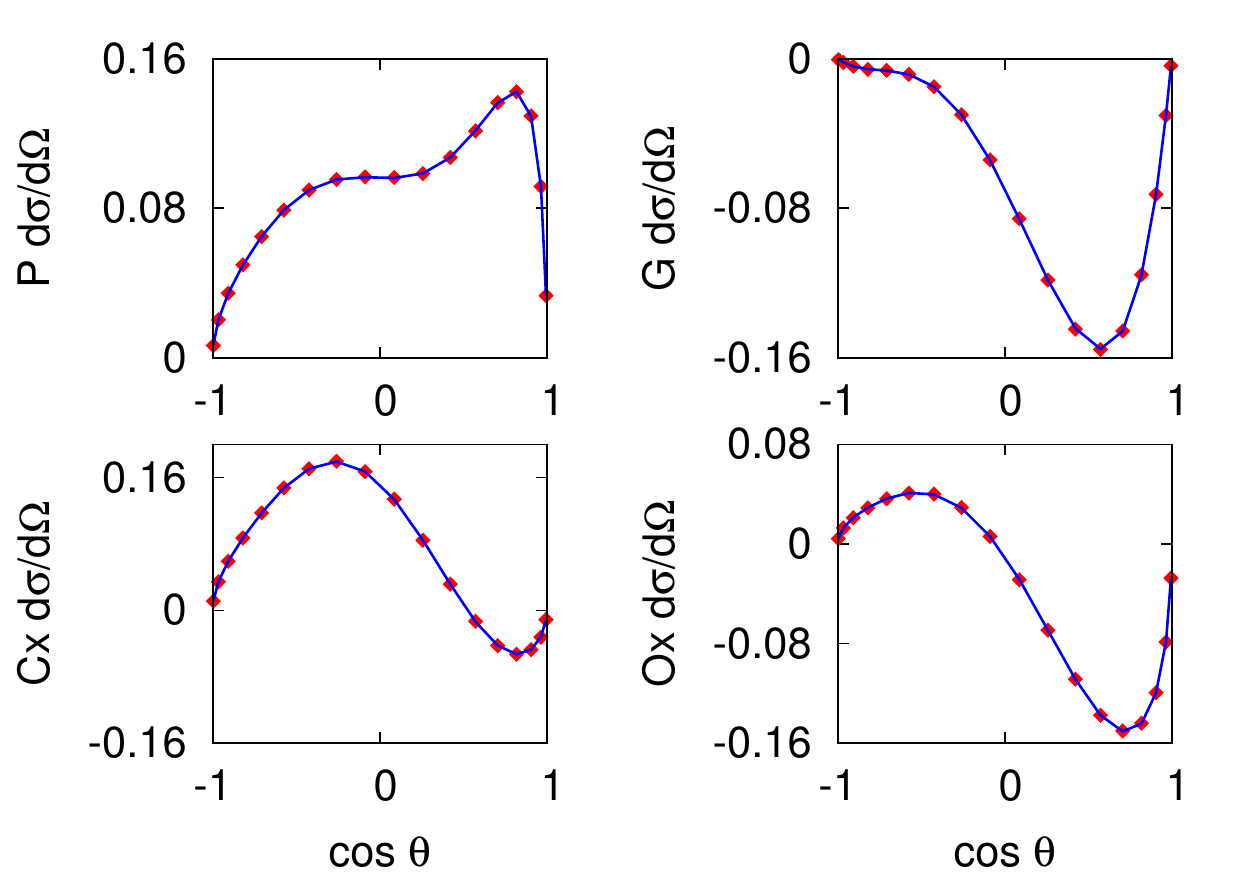}
\end{center}
\vspace*{-0.5cm}
\caption{\label{Observables}(Color online) Complete set of observables for $\eta$ photo-production given in the form of numeric data created at $W= 1769.8$ MeV and at 18 angles using the ETA-MAID15a model (red symbols) and a typical SE fit to the data (full line). }
\end{figure}

 In Fig.~\ref{Observables} we show a complete set of numeric data created at 18 angles (red symbols), and the typical SE fit (full line) at one representative energy of $W = 1769.80$ MeV. All SE fits are similar, and among themselves indistinguishable on the given scale.

\begin{figure}[h!]
\begin{center}
\includegraphics[width=0.7\textwidth]{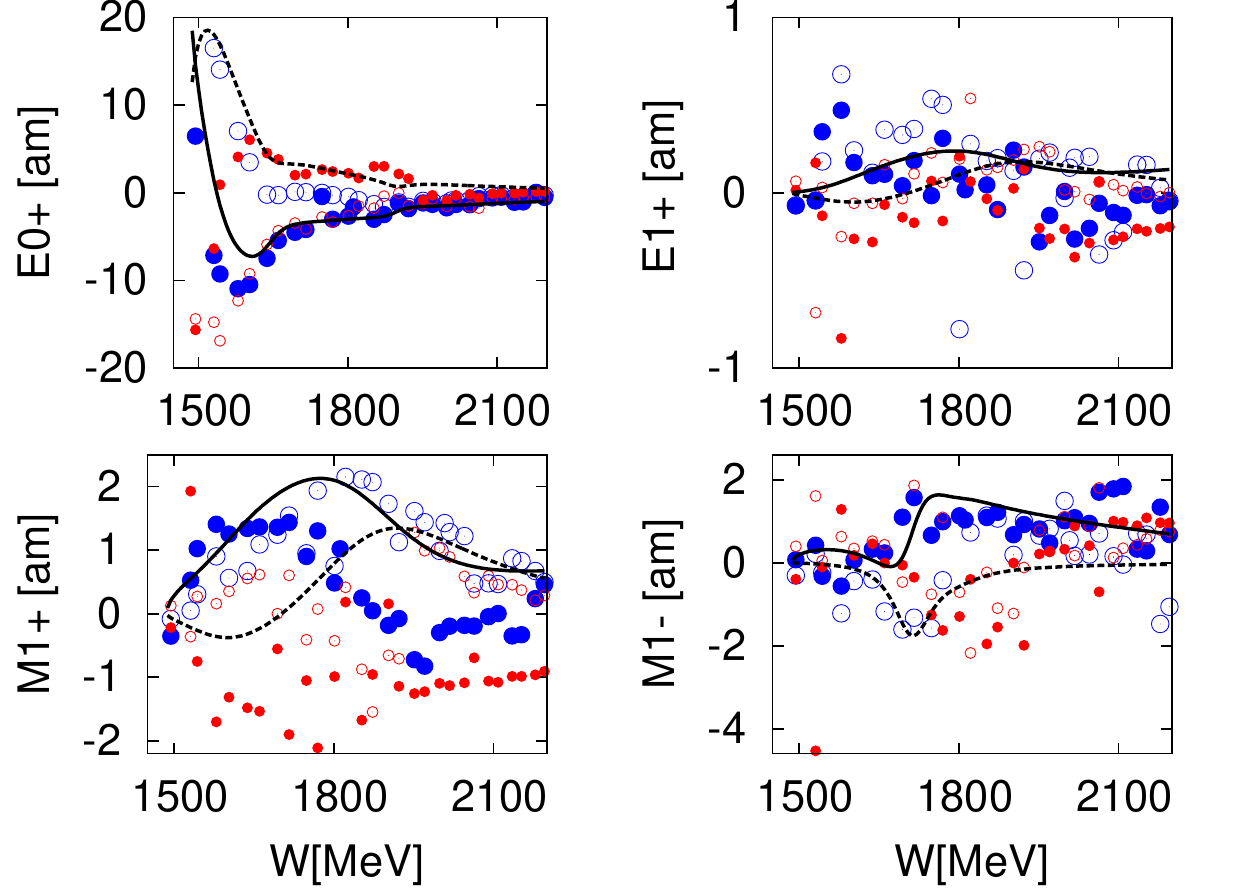}
\end{center}
\vspace*{-0.5cm}
\caption{\label{Fig3}(Color online) Plots of the $E_{0+}$, $M_{1-}$, $E_{1+}$, and $M_{1+}$
multipoles. Full and dashed black lines give the real and imaginary part of the ETA-MAID15a generating model.
Discrete blue (bigger) and red (smaller) symbols are obtained  with the unconstrained, $L_{max}=5$  fits of a complete set of
observables  generated as numeric data from the ETA-MAID15a model of ref. \cite{MAID15a},
with the initial fitting values taken from the ETA-MAID16a \cite{MAID16a} and the
Bonn-Gatchina \cite{BoGaweb} models respectively.
Filled symbols represent the real parts and open symbols give the  imaginary parts. \\  Attometer(am) $\equiv$ milli-fermi(m\,fm).}
\end{figure}

In Fig.~\ref{Fig3} we show an example of three very different sets of SE multipoles which fit the complete numeric data set equally well to a high precision.

\begin{figure}[h!] \hspace*{-0.5cm}
\begin{center}
\includegraphics[width=0.4\textwidth]{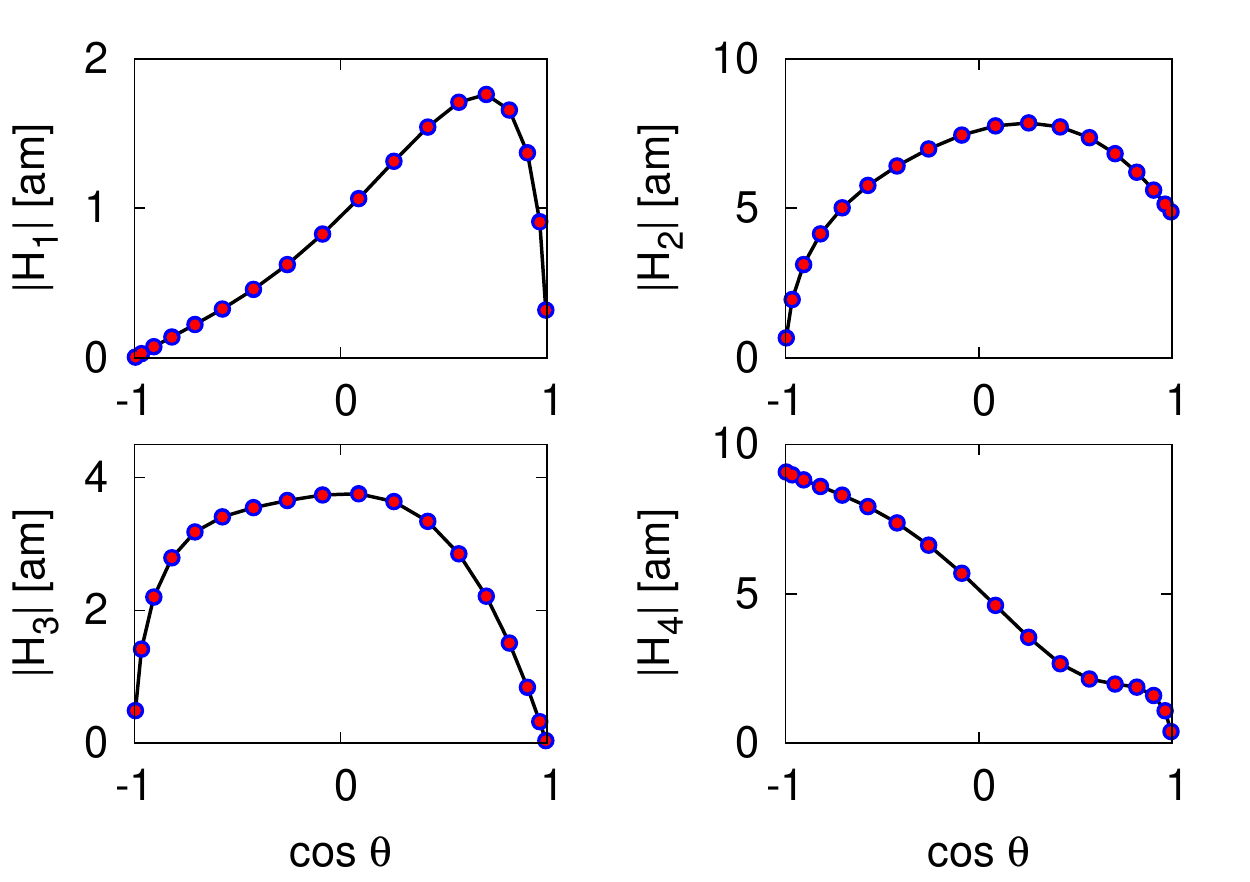} \hspace*{1.cm}
\includegraphics[width=0.4\textwidth]{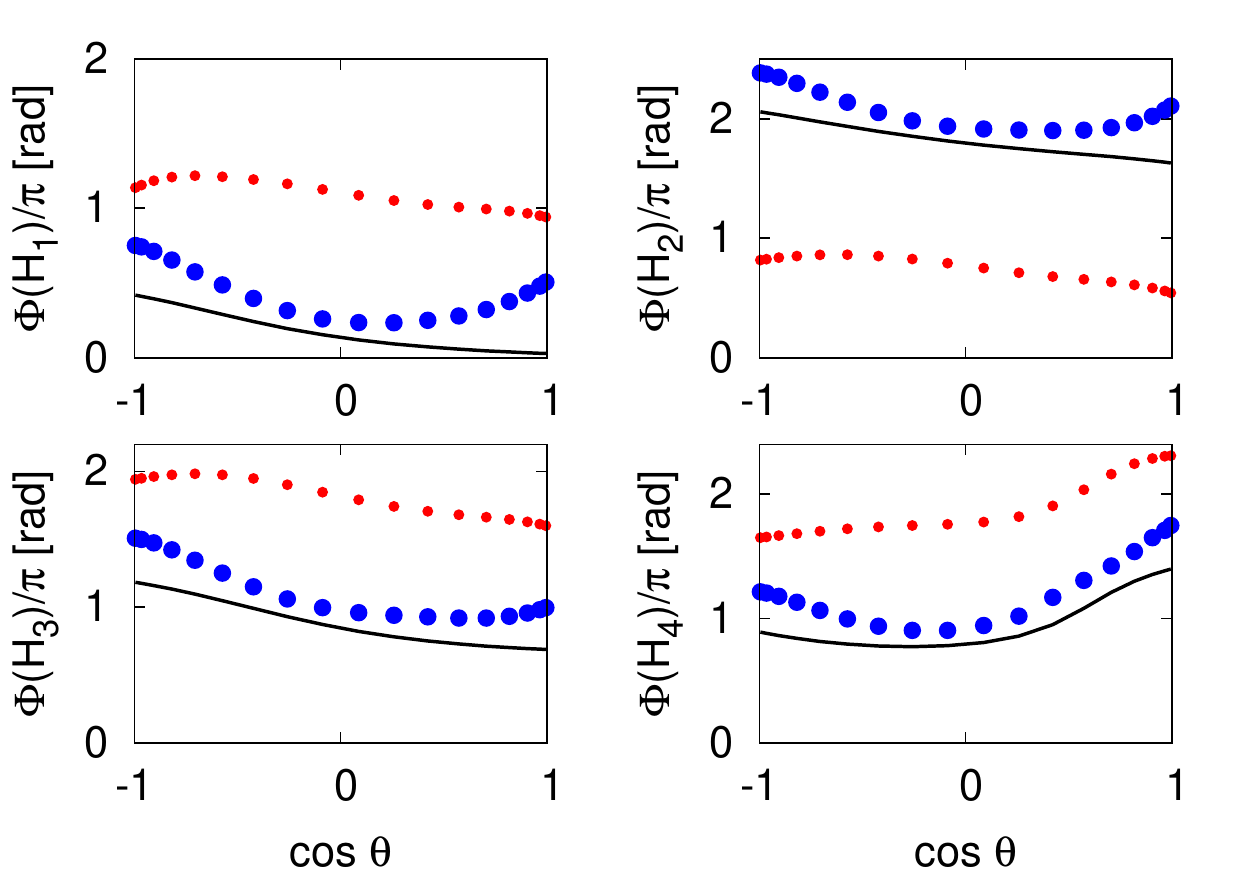}   \\
\caption{\label{Fig4}(Color online)  We show three sets of helicity amplitudes for all three sets of multipoles at an randomly chosen energy  $W= 1660.4$ MeV. Fig (a) shows absolute values, and Fig (b) phases of all four helicity amplitudes.  \\
 The figure coding is the same as in Fig.~\ref{Fig3}.  Attometer(am) $\equiv$ milli-fermi(m\,fm). }
\end{center}
\end{figure}

\begin{figure}[h!] \hspace*{-0.5cm}
\begin{center}
\includegraphics[width=0.4\textwidth]{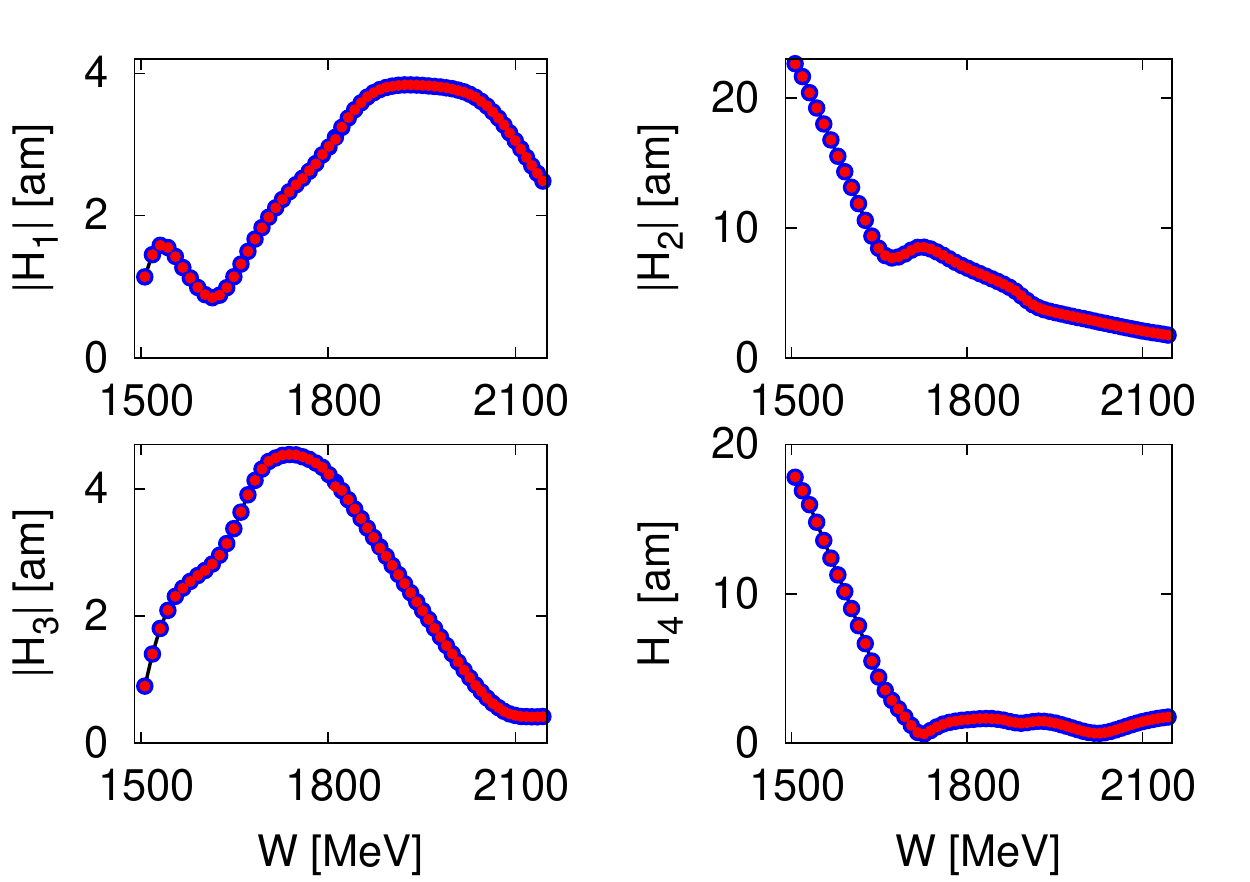} \hspace*{1.cm}
\includegraphics[width=0.4\textwidth]{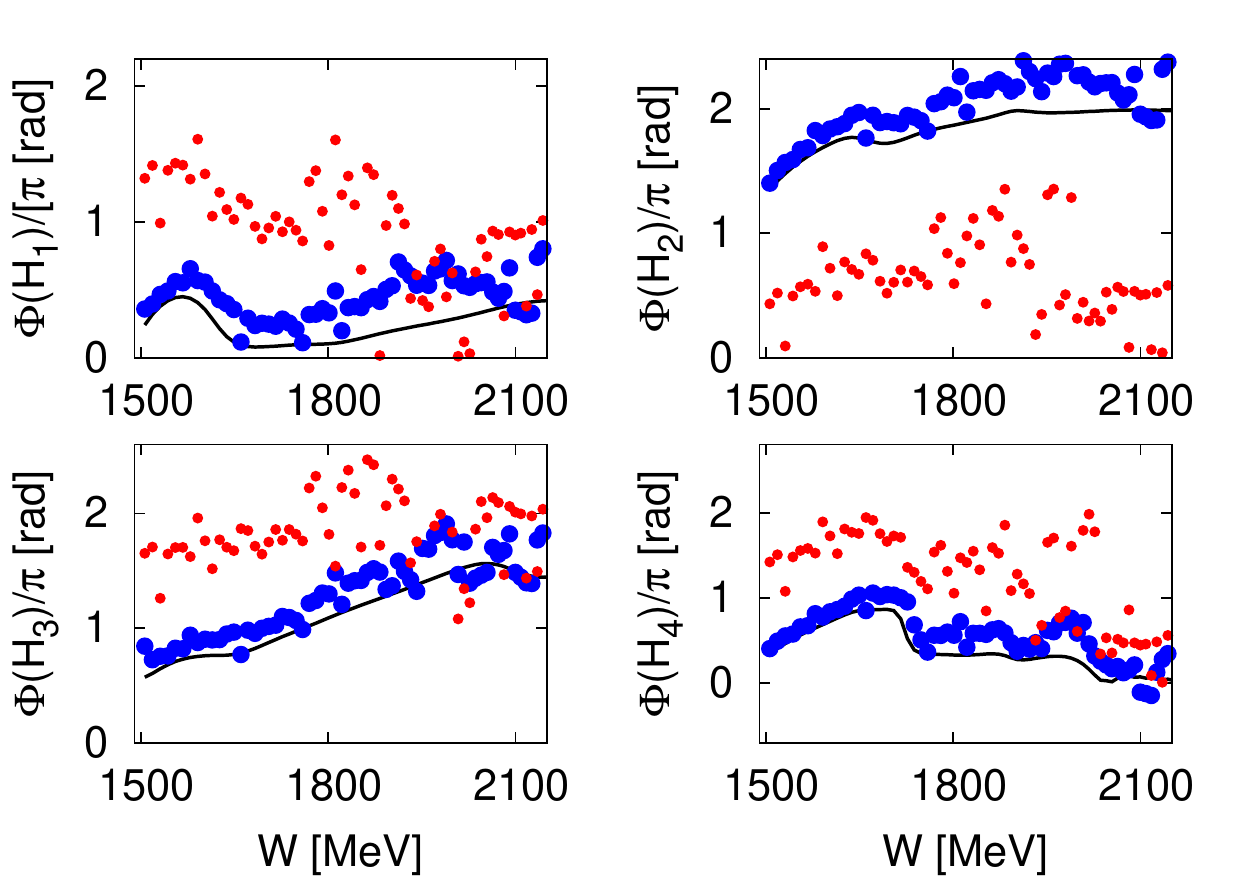}   \\
\caption{\label{Fig5}(Color online)  Excitation curves of all
helicity amplitudes, for all three sets of multipoles, at an randomly
chosen value of  $\cos \theta= 0.2588$.  Fig (a) shows absolute values, and Fig (b) phases of all four helicity amplitudes. \\
 The figure coding is the same as in Fig.~\ref{Fig3}.    Attometer(am) $\equiv$ milli-fermi(m\,fm). }
\end{center}
\end{figure}

The two sets are discrete and are obtained in the unconstrained SE fit by setting the initial fitting values first to
the ETA-MAID16a \cite{MAID16a} (SE$^{16a}$)  and second to
Bonn-Gatchina \cite{BoGaweb} (SE$^{BG}$)
model values (blue and red symbols respectively). Observe that both discrete sets are discontinuous, and different.  The third set of multipoles are the generating ETA-MAID15a model  \cite{MAID15a} input which are displayed as full and  dashed black continuous lines. This means that we have two issues to resolve: the first issue is discontinuity, and the second one is non-uniqueness.
\\ \\  \newpage
This results is well known but discouraging. Obtained unconstrained sets of SE multipoles where only two out of many possible solutions are given in Fig.~\ref{Fig3}, are useless for any sensible analysis. Given the problems with uniqueness and continuity, there is no reasonable way to assign poles, based on these sets of multipoles.
\\ \\\noindent
 It is the aim of this paper to remedy this, and let us first attack the discontinuity problem.
\\ \\
 As a first step, we consider helicity amplitudes which can be obtained
directly from the data. Here we have obtained these amplitudes from
a complete sete of numeric data.
In Fig.~\ref{Fig4}, we show absolute values and phases of helicity amplitudes
corresponding to all three sets of multipoles from Fig.~\ref{Fig3}
at an randomly chosen energy \mbox{$W= 1660.4$~MeV}, and in  Fig.~\ref{Fig5} an excitation curve of all four helicity amplitudes
at an randomly chosen angle  $\cos \theta=0.2588$.
\\ \\ \noindent
From Figs.~\ref{Fig4} and ~\ref{Fig5} we see:
\begin{enumerate}
        \item Absolute values for all solutions are always identical and equal to absolute values of the generating solution MAID15a.  This is consistent with the fact that the continuum ambiguity, as the only source of non-uniqueness, influences only the phase and not the modulus of any helicity amplitude.
        \item The phases of all three solutions are different for the cases of a fixed energy and a fixed angle. They are continuous in first case and discontinuous in the second.
       \item  The phases at a fixed energy (Fig.~\ref{Fig4} are forced to be continuous by modeling the angle dependence via partial-wave decomposition,  but they are different as they are generated by different models. Energy dependent, generating model Eta-MAID15a has its own phase, and both SE solutions retain the phase of the initial solution used in the fit (blue dots the phase of ETA-MAID16a model, and red dots the phase of Bon-Gatchina model). However, for the fixed angle (Fig.~\ref{Fig5}),  the discontinuity in phases in helicity-amplitudes excitation functions of discrete solutions appear because no continuity condition like the one in the former case for angles at a fixed energy is imposed between neighbouring energy points at a fixed angle. At each energy the fit starts from the continuous initial solution, but the fit randomly chooses its phase out of infinite number of phase equivalent solutions at a given energy. So, this is a machine-generated effect.

\end{enumerate}

\begin{figure}[h!] \hspace*{-0.5cm}
\begin{center}
\includegraphics[width=0.4\textwidth]{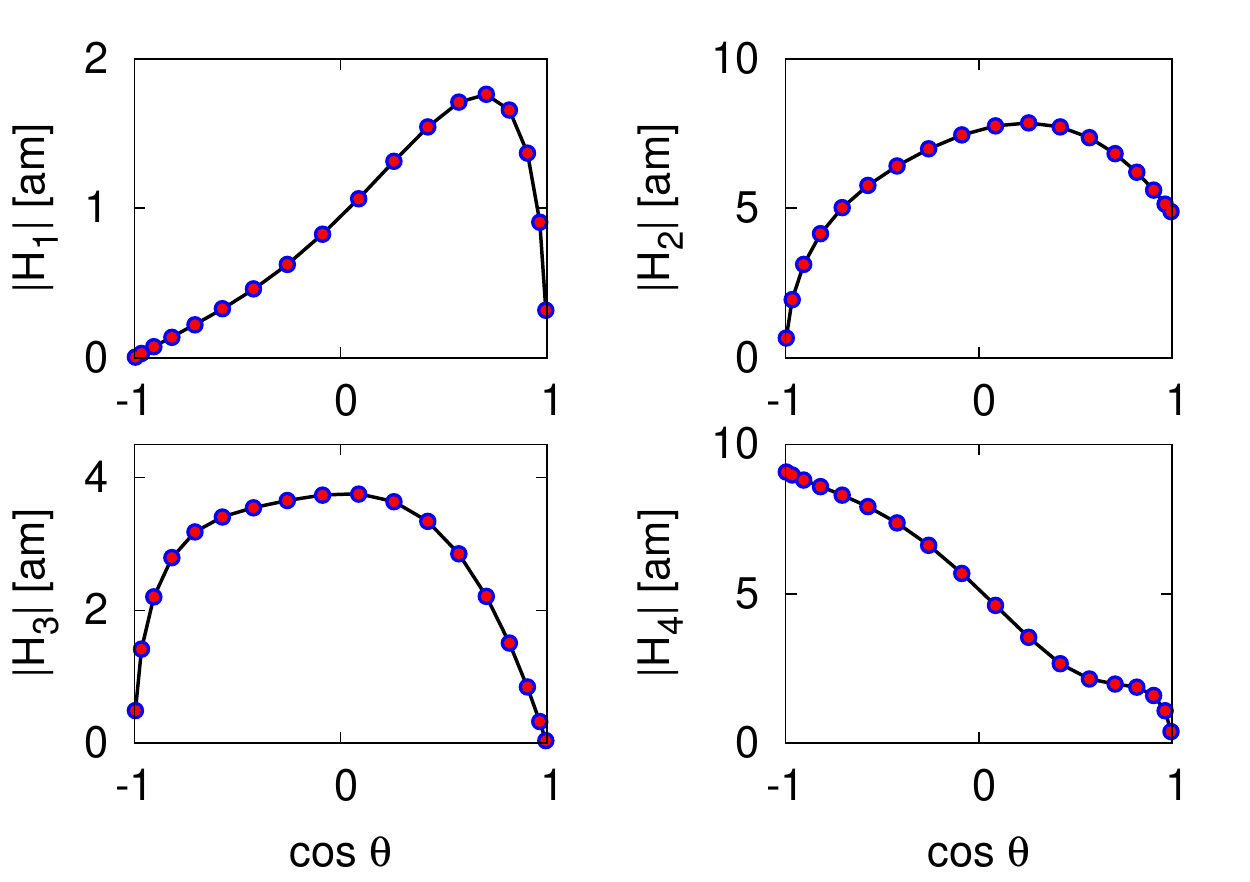} \hspace*{1.cm}
\includegraphics[width=0.4\textwidth]{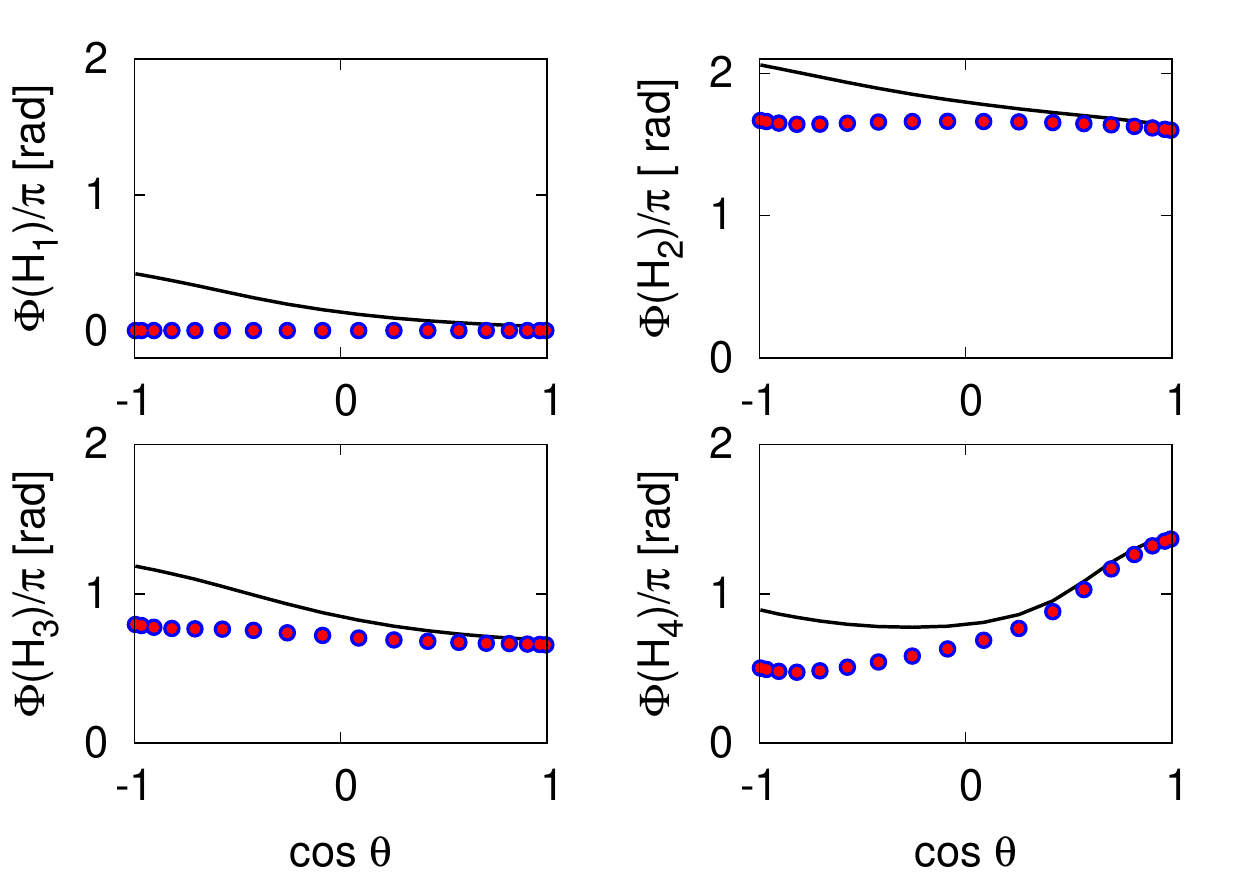}   \\
\caption{\label{Fig6}(Color online)  We show two sets of rotated helicity amplitudes, and a generating model value for all multipoles of Fig.~\ref{Fig3} at an randomly chosen energy  $W= 1660.4$ MeV. Rotation was performed with  $\mathcal{\psi}(W,\theta) = 0$.   Fig (a) shows absolute values, and Fig (b) phases of all four helicity amplitudes.
 \\ The figure coding is the same as in Fig.~\ref{Fig3}.   Attometer(am) $\equiv$ milli-fermi(m\,fm). }
\end{center}
\end{figure}

\begin{figure}[h!] \hspace*{-0.5cm}
\begin{center}
\includegraphics[width=0.4\textwidth]{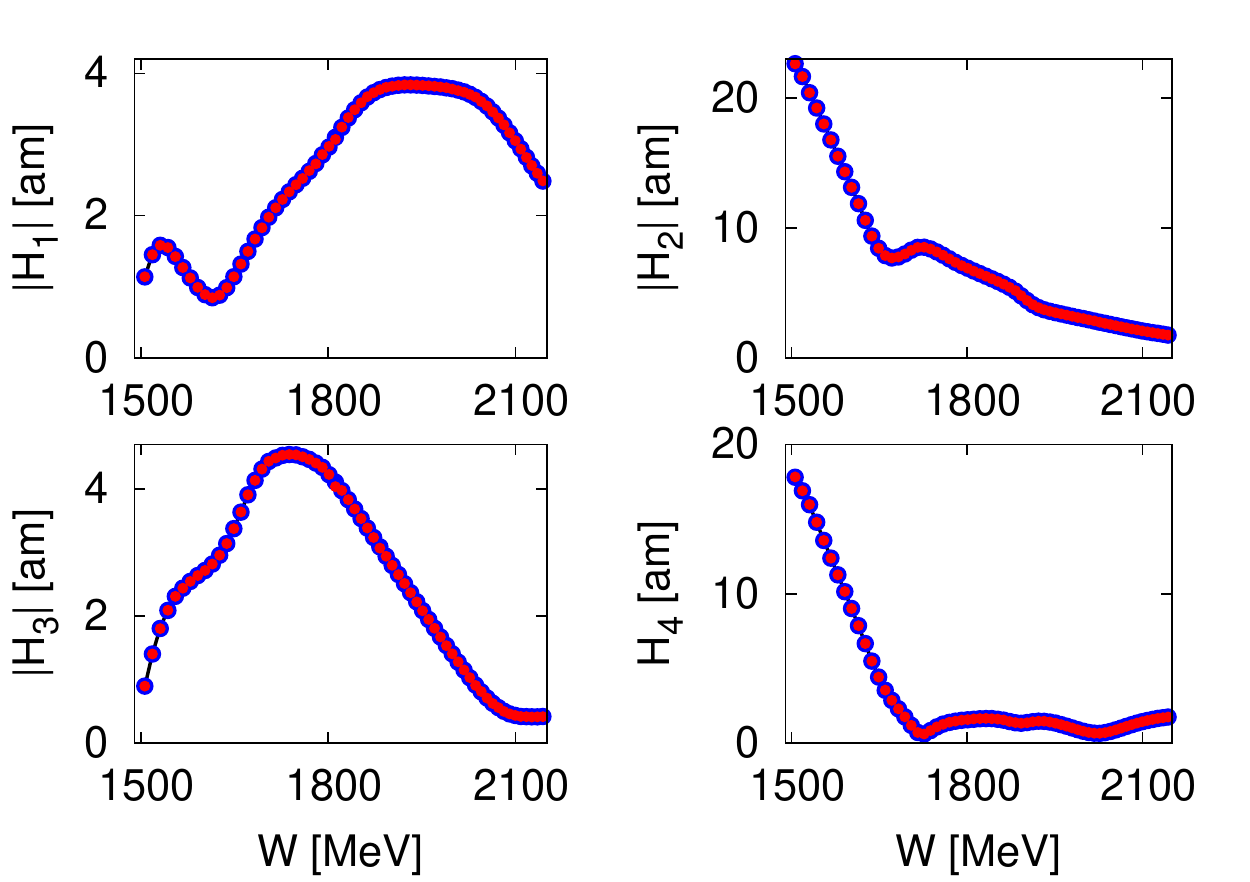} \hspace*{1.cm}
\includegraphics[width=0.4\textwidth]{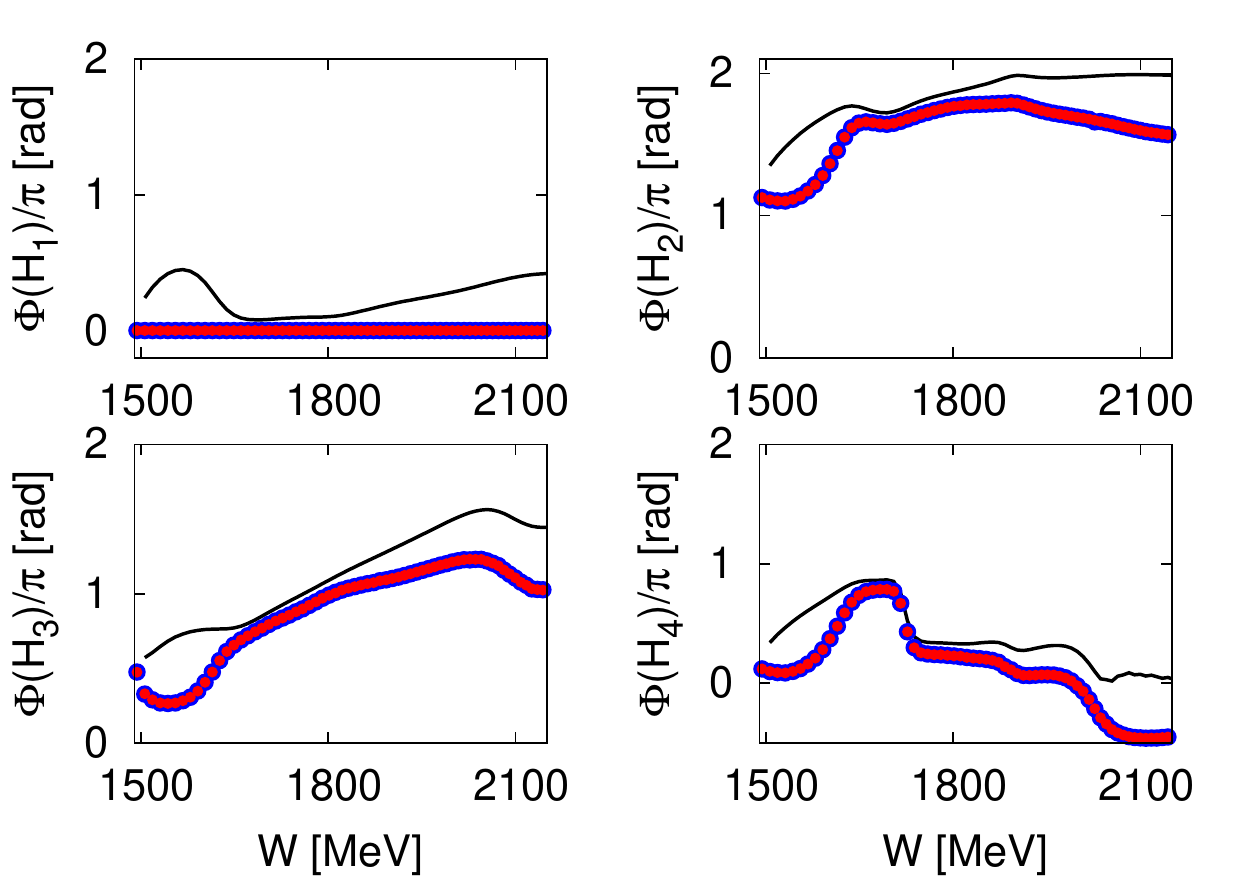}   \\
\caption{\label{Fig7}(Color online)  Excitation curves of two rotated helicity amplitudes and a generating model one for all three sets of multipoles of Fig.~\ref{Fig3}, at an randomly chosen value of  $\cos \theta= 0.2588$.  Rotation was performed with  $\mathcal{\psi}(W,\theta) = 0$.  Fig (a) shows absolute values, and Fig (b) phases of all four helicity amplitudes. \\ The figure coding is the same as in Fig.~\ref{Fig3}.  Attometer(am) $\equiv$ milli-fermi(m\,fm).  }
\end{center}
\end{figure}

First let us show that the source for discontinuity of unconstrained SE multipoles (Fig.~\ref{Fig3}) lie in the discontinuity of the excitation function of phases.
\\ \\ \indent
As from Eq.~(1) we know that a complete set of observables is invariant with respect to an arbitrary phase rotation, we are free to introduce any rotation we choose, provided that it does not violate basic physics principles like inelastic unitarity for instance. So, we propose to introduce an angle-dependent phase rotation simultaneously for all four helicity amplitudes in the following way:
\be \label{Phaserotation}
\tilde{H}_k^{SE}(W, \theta) &=& H_k^{SE}(W, \theta) \, \cdot \, e^{\di \, \mathcal{\psi}(W,\theta) - \, \di \, \Phi_{H_1}^{SE}(W, \theta)} \nonumber  \\
k  & = & 1,\ldots,4
\ee
where $\mathcal{\psi}(W,\theta)$ is the arbitrary, but in all variables continuous phase, including the constant.  Introducing this invariant phase rotation simultaneously for all four helicity amplitudes we explicitly eliminate the discontinuity in excitation curve for helicity amplitude phase $\Phi_{H_1}^{SE}(W, \theta)$.  However, it also automatically eliminates discontinuities in the excitation curves of remaining three helicity-amplitude phases, because it is enough to fix one helicity-amplitude phase to obtain one and the same phase-equivalent solutions at all analyzed energies (all helicity amplitudes become continuous).
\newpage
In Figs.~\ref{Fig6} and \ref{Fig7}  we show absolute values and phases of all four helicity amplitudes at an randomly chosen energy of  $W= 1660.4$ MeV and  angle  $\cos \theta=0.2588$  for the simplest phase rotation $\mathcal{\psi}(W,\theta) = 0$  which entirely eliminates the imaginary  part of $H_1(W, \theta)$\footnote{ Let us just mention that the phase of any out of four helicity amplitudes may have been used. }. In this way we have obtained a non-unique, but continuous and hence useful solution.


Observe that phase $\Phi_{H_1}^{SE}(W, \theta)$  is always zero, but phases for other three helicity amplitudes are now continuous.
\newpage
When we make a partial-wave decomposition of these amplitudes we obtain a set of continuous multipoles at a new phase generated by Eq.~(\ref{Phaserotation}) at $\mathcal{\psi}(W,\theta) = 0$. We show again only the lowest multipoles in Fig.~\ref{Figrotmult}:
 \begin{figure}[h!]
\begin{center}
\includegraphics[width=0.75\textwidth]{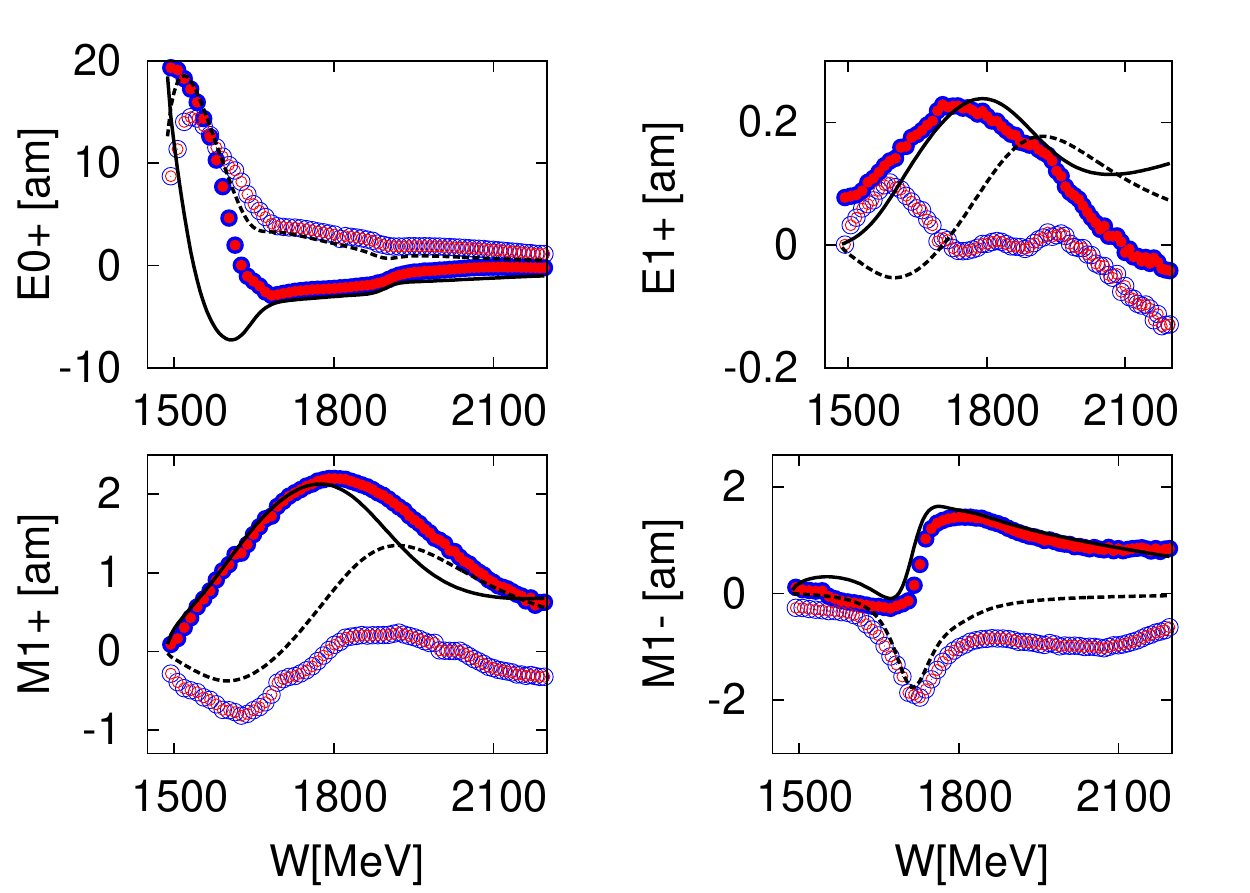}
\end{center}
\vspace*{-0.5cm}
\caption{\label{Figrotmult}(Color online) Plots of the $E_{0+}$, $M_{1-}$, $E_{1+}$, and $M_{1+}$
multipoles obtained after the phase rotation $\mathcal{\psi}(W,\theta) = 0$.   The figure coding is the same as in Fig.~\ref{Fig3}.  Attometer(am) $\equiv$ milli-fermi(m\,fm).
 }
\end{figure}

As we said before, unconstrained SE multipoles obtained in Fig.~\ref{Fig3} are useless. Contrary to that, and in spite of not being unique as they depend on the choice of the arbitrary phase rotation phase $\mathcal{\psi}(W,\theta)$, multipoles obtained via Eq.~(\ref{Phaserotation}) are continuous. As an example, the lowest multipoles are for the simplest phase  $\mathcal{\psi}(W,\theta) = 0$  given in  Fig.~\ref{Figrotmult}.  We see that  all multipoles  can now be analyzed;  analytic structure - poles in particular, can be found using L+P formalism, and all poles in all partial waves for that particular phase may be established. {But then, as it directly follows from Eq.~(\ref{Eq:mixing-expanded}), all possible poles in all partial waves are also found for any other (also the true)  phase as the phase rotations only reshuffle once established poles throughout all partial waves.  This means that a pole due to cancelations may after the rotation disappear  in a certain partial wave, but it will certainly reappear somewhere else, and as a rule not in only one partial wave.  Hence, even when the poles are reliably established, their  quantum numbers are not; for that we do need a knowledge of the true phase.} So, by making the phase continuous using Eq.~(\ref{Phaserotation}),  the reaction can be analyzed, some conclusions on the resonance content of the process can be drawn, but the final answer will not be given until the final phase is determined.
 Of course, taking the constant phase is the simplest way out. However, when we analyze real data, we shall also always have a phase from some theoretical model which is correctly describing the analyzed data base at our disposal, so instead of taking $\mathcal{\psi}(W,\theta) = 0.$  we shall be able to take $\mathcal{\psi}(W,\theta) = \Phi_{th}(W, \theta)$.  However, reasoning will remain the same as for the constant phase, and we do not see any real advantage if we use the theoretical phase instead.

At this moment let us point out that this discussion is far from being complete.
  There is a number of possible pathological situations like { vanishing or small $L_0$} or the possibility pointed out by G. H\"ohler in \cite{Hohref} that at two energies (1680 and 1880 MeV) in the pion-nucleon scattering system there are resonances in all partial waves (up to a certain value of L).
   These are valid points to be extensively discussed when the final criterion for determining poles and their true quantum numbers is considered and a scheme is proposed. However, in the present work, we have outlined new insights into the implications of an unknown phase for partial-wave analysis; an understanding of relative phases in terms of dynamics is beyond the scope of this study.
 We believe that it is important for the reader to understand and accept the facts that partial wave decomposition is non-unique, and that it depends on the choice of the phase undetermined directly by experiment.  Possible criteria and procedures  how this can be overcome is another and complicated issue, and will be the topic of further studies.

\section{Conclusions}
Let us summarize by affirming that taking into consideration the angle dependent continuum ambiguity phase is not only an academic issue; it is essential. The problem it introduces is that it mixes partial waves and hence modifies resonance quantum number association. On the other hand, employing this phase-rotation freedom enables one to obtain an up-to-a phase unique solution in the unconstrained PWA of a complete set of observables.

However, in practical, day-to-day calculations the continuum ambiguity must, in some way, be taken into account to actually
obtain a continuous partial-wave solution. Without fixing the overall angle and energy dependent phase,
a whole class of equivalent reaction amplitude sets, connected via rotations, exist and that produces discontinuities in unconstrained  SE PWA.  We eliminate this problem by fixing the unknown phase to any continuous one. In this way a solution with a continuous, but predefined phase is obtained, and a pole search in now continuous partial waves can be undertaken. Let us point out that it  by no means represents the loss of generality.
Namely, once we have obtained a unique, continuous solution with a chosen phase, we can
obtain all poles of the system by analyzing analytic structure of partial waves for this particular phase. We claim that it is enough to make discontinuous partial waves continuous at any phase to be able to find all poles of the reaction in all partial waves. However, their quantum numbers will be ill defined without knowing the true phase.  Namely, in spite of being mixed throughout partial waves, all poles from the original phase survive, and are contained in each solution with any given phase. This can be seen by analysing Eq.~(\ref{Eq:mixing-expanded}). We know that the true phase exists, and that for this phase each partial wave contains $n_l$ poles, so if the whole process is described with L partial waves we have all in all $N = \sum_{l=0}^{L} n_l$ different poles with good quantum numbers. This means that each pole is contained in one and only one partial wave. However, from this phase we can reach any other phase by a phase rotation. This phase will redistribute partial waves on the basis of Eq.~(\ref{Eq:mixing-expanded}). In that phase each partial wave may have more that $n_l$ poles (some of them are admixed from other partial waves because of the rotation), and the whole system will have more than $N$ poles. However, some poles will be found multiple number of times, in different partial waves, including the true partial wave. Therefore, when we analyze partial waves at any phase, and when we find a pole, we know that we have found a resonance, but its quantum numbers are not well defined.

Let us finally state that the continuum ambiguity phase is not just a mathematical fiction without possibility to be determined. First, it definitely exists, and seriously influences our reasoning. Second, it can also be  determined; definitely in coupled-channel calculations { as this is a way of satisfying unitarity. Each time a new channel is added the unitary sum is modified. The coupled-channel method guaranties that the system is always unitary.}  When only one channel is analyzed, unitarity constraints relating real and imaginary parts of the partial-wave T-matrices are lost after the first inelastic threshold opens so continuum ambiguities arises, but it is automatically restored when all channels are included in coupled-channel calculations (uncontrolled loss of flux in one channel is controlled by including all of them). So, coupled-channel formalisms are the natural way to eliminate continuum ambiguities.
 However, let us keep in mind that existing multi-channel analyses are also
only approximations - one has to use a model and can't include all possible channels.
Therefore, the true phase could only be determined approximately by coupled channels and would have a model dependence.

\section*{Acknowledgements}
This work was supported by the Deutsche Forschungsgemeinschaft (SFB TR16 and 1044). The work of
RW was supported by the U.S. Department of Energy grant DE-SC0016582.

\end{document}